\documentclass[aps,prl,twocolumn,showpacs,preprintnumbers,amsmath,amssymb]{revtex4}
\usepackage{graphicx}
\usepackage{dcolumn}
\usepackage{mathrsfs}
\usepackage{bm}

\usepackage{amsfonts}
\usepackage{amssymb}
\usepackage{amsmath}
\usepackage{color}

\begin{document}

\preprint{}

\title{Polarization Entangled Photons at X-Ray Energies}

\author{S. Shwartz }\email{shwartz@stanford.edu} \author{ S. E. Harris}

\address{
Edward L. Ginzton Laboratory, Stanford University, Stanford,
California 94305}
\date{\today}

\begin{abstract}
We show that polarization entangled photons at x-ray energies can be generated via spontaneous parametric down conversion.  Each of the four Bell states can be generated by choosing the angle of incidence and polarization of the pumping beam.
\end{abstract}


\maketitle

Spontaneous parametric down-conversion at hard x-ray energies was first proposed by Freund and  Levine in 1969 \cite{Freund}, and first demonstrated experimentally by Eisenberger and McCall about two years latter \cite{Eisenberger}. In that work a hard x-ray tube was used as the pump, and coincidence counts at the signal and idler were measured at the rate of a few counts per hour. The first experiment using a synchrotron was done in 1997 by Yoda et al.  \cite{Yoda} with a counting rate of  6 counts per hour. During the years 1998-2003, Adams and collaborators conducted a series of experiments  and improved the coincidence count rate to about 1 count per 13 seconds \cite{Adams}. Recent and expected improvements in brilliance and beam quality of synchrotron x-ray sources, together with new facilities such as the x-ray free electron laser and energy recovering linacs  \cite{Adams}, offer the possibility of extending the concepts of quantum optics as developed in the visible portion of the electromagnetic spectrum  \cite{Bouwmeester}  to x-ray wavelengths.

As a step toward this extension, this Letter describes a method for generating polarization entangled photons, in pure Bell states, at x-ray wavelengths. The technique is straight-forward and makes use of the selection rules that are associated with a phase matched plasma-like x-ray nonlinearity \cite{Adams}. In the following paragraphs we will show that by choosing the polarization and angle of incidence of the pumping beam, and working off of the degenerate frequency, that each of the four Bell states may be generated. A consequence of  this work is that, in each of the previous x-ray down conversion  experiments mentioned above, the generated photon pairs were polarization entangled, but in no case were they in a pure Bell's state.

Before proceeding we note two previous suggestions for generating entangled photons at x-ray energies. The first is a proposal by Sch\"{u}tzhold et al. who have suggested the use of ultra-relativistic electrons accelerated by a strong periodic electromagnetic field (for example, a laser or undulator) to create entangled photon pairs in the multi-keV regime \cite{Schutzhold}. The second is a proposal by P\`{a}lffy et al. who suggest generating single-photon entangles states by control of nuclear forward scattering \cite{Palffy}.

We start by discussing the nonlinearity. The central concept of the nonlinearity at x-ray energies is that, since x-ray photons have energies that are large as compared to the electron binding energy of low-Z atoms, that the x-ray nonlinearity of an element such as diamond may be calculated by treating all of the electrons in the atom as free particles, and therefore treating the nonlinear medium as a very dense cold plasma \cite{Freund,Eisenberger}. This x-ray nonlinearity is of second order so that two frequencies may add or subtract to generate a third frequency. Three processes that are at first glance seemingly different contribute to the x-ray nonlinearity. These are: 1) A Lorentz force term where the electron velocity caused by an incident electric field at frequency $\omega_{1}$ mixes with the magnetic field of frequency $\omega_{2}$ to generate a force and current at frequency $\omega_{3}=\omega_{1}+\omega_{2}$. 2)  A term that depends on the spatial variation of charge density, and 3) A term that depends on the spatial variation of velocity. Each of these processes produces a driving current with a k-vector that is the sum of the k vectors of the applied fields and of the lattice.

In order to satisfy permutation symmetry and to conserve photons in a three-frequency nonlinear optical process, it is essential that the three processes of the previous paragraph all be retained in the calculation of the nonlinearity  \cite{Shen}. It is easily shown that each of the above processes,  for example the Lorentz force process, does not in its own right satisfy permutation symmetry.

\begin{figure}[t]
\begin{center}
\includegraphics*[width=2.5 truein]{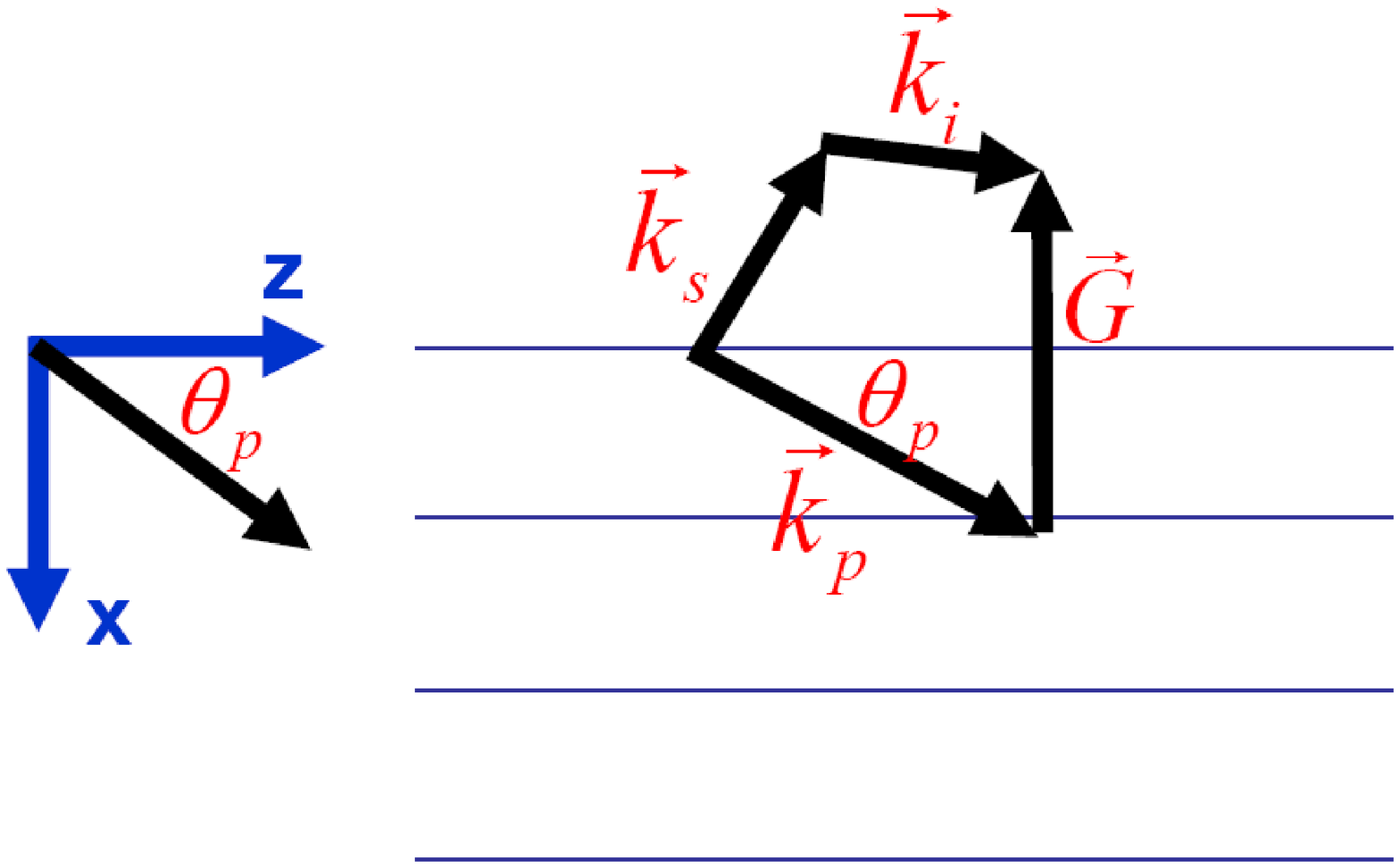}
\end{center}
\caption{Phase matching diagram for x-ray spontaneous down conversion .  $\vec{k}_{s}$, $\vec{k}_{i}$,  and $\vec{k}_{p}$ are the wave vectors of the signal, idler, and pump fields.  $\vec{G}$ is the reciprocal lattice vector of the diamond crystal.}
\label{fig:coordinate}
\end{figure}

A typical phase matching diagram for parametric down conversion is shown in Fig. ~\ref{fig:coordinate}. With the k-vectors  $\vec{k}_{s}$,
$\vec{k}_{i}$, $\vec{k}_{p}$ denoting the k-vectors of the signal, idler, pump fields and $\vec{G}$ denoting the reciprocal lattice vector, the phase matching condition for parametric down conversion is $\vec{k}_{s}+\vec{k}_{i}=\vec{k}_{p}+\vec{G}$. With the unit vectors  $\hat{e}_{j}$ denoting the polarization of the respective electric fields, these fields are written as
$ \vec{E}_{j}(\vec{r},t)= \frac{E_{j}}{2}\exp [- i(\omega_{j} t-\vec{k}_{j}\cdot\vec{r})] \hat{e}_{j}+c.c.$  Working in the cold plasma approximation, we perturbatively calculate the nonlinear current density $\vec{J}_{s}(\vec{r},t)=\rho_{s}(\vec{r},t)\vec{v}_{s}(\vec{r},t)$ at the signal frequency \cite{Eisenberger,Jha,Nazarkin}
 to obtain
\begin{align}
\label{eq:current2}
   \vec{J}_{s}(\vec{r},t)&=i\frac{q^{2}}{m^{2}}\Big[-\frac{\rho_{0}(\vec{r})}{\omega_{s}\omega_{i}\omega_{p}}\nabla\Big(\vec{E}_{p}(\vec{r},t)\cdot\vec{E}_{i}(\vec{r},t)\Big)\nonumber\\
   &-\frac{1}{\omega_{p}^{2}\omega_{i}}\Big(\nabla\rho_{0}(\vec{r})\cdot\vec{E}_{p}(\vec{r},t)\Big)\vec{E}_{i}(\vec{r},t)\nonumber\\
   &+\frac{1}{\omega_{i}^{2}\omega_{p}}\Big(\nabla\rho_{0}(\vec{r})\cdot \vec{E}_{i}(\vec{r},t)\Big)\vec{E}_{p}(\vec{r},t)\Big]
   \end{align}
Here  q and m are the electron charge and mass, and $\rho_{0}(\vec{r})$ is the electron density in the absence of the pumping beam. We substitute the expressions for the electric fields into Eq.~\eqref{eq:current2} and project the nonlinear current density against the direction of the signal electric field. We assume phase matching \cite{Freund,Nazarkin} with  the reciprocal lattice vector $\vec{G}$ so that the unperturbed electron charge density is taken as  $\rho_{0}(\vec{r})=\rho_{g}\exp[i\vec{G}\cdot\vec{r}]$. The envelope of the nonlinear current density is then
\begin{align}
\label{eq:permutation2}
   J_{s}&=-\frac{q^{2}\rho_{g}\omega_{s}E_{p}E^{\ast}_{i}}{4m^{2}\omega_{p}^{2}\omega_{i}^{2}\omega_{s}^{2}}\Big[\omega_{i}\omega_{p}\big(\vec{G}\cdot \hat{e}_{s}\big)\big(\hat{e}_{p}\cdot\hat{e}_{i}\big)\nonumber\\
   &-\omega_{s}\omega_{i}\big(\vec{G}\cdot\hat{e}_{p}\big)\big(\hat{e}_{i}\cdot\hat{e}_{s}\big)+\omega_{s}\omega_{p}\big(\vec{G}\cdot \hat{e}_{i}\big)\big(\hat{e}_{p}\cdot\hat{e}_{s}\big)\Big]
   \end{align}
As shown in Fig.~\ref{fig:coordinate} we define the scattering plane as the plane containing the k-vector of the pumping beam and the lattice k-vector, and assume that the k-vectors of the signal and idler beams are also in this plane. From  Eq.~\eqref{eq:permutation2} we find the following selection rules \cite{Adams}: (1) If the polarization of the pump is in the scattering plane, the polarizations of the signal and the idler photons must both be either in the scattering plane or must both be normal to the scattering plane. (2) If the polarization of the pump is normal to the scattering plane, then either the signal polarization is in the scattering plane and the idler polarization is normal to the scattering plane or vise-versa. Polarization entanglement requires that the polarization of the idler is uniquely determined by the polarization of the signal.  For the pump polarized either in, or orthogonal to, the scattering plane the down converted signal and idler photons are therefore entangled.

The polarization of the entangled photon pairs is determined by the driving current as described by Eq.~(2) and is not influenced by the temporal or angular dispersion of the system.  To calculate Glauber correlations and the biphoton generation rate, we work in the Heisenberg picture and write a pair coupled equations for each of the biphoton pairs  \cite{Gatti}. For example, for the entangled state $| \phi>=(|H_{s},V_{i}>+|V_{s},H_{i}>)/\sqrt{2}$, we write a pair of coupled equations for the state $|H_{s},V_{i}>$ and a second set of coupled equations for the state
 $|V_{s},H_{i}>$. It is critical that the dependence on frequency and angle of emission of the k-vector mismatch function is  the same in each pair of coupled equations. Each of the biphoton wave packets $|H_{s},V_{i}>$ , or  $|V_{s},H_{i}>$ has significant dispersion and the temporal and spatial correlations between these packets will vary with position and angle. But because the k-vector mismatch is independent of polarization and is the same for each packet, the polarization correlations are determined by the selection rules associated with Eq.~(2), and do not vary with propagation.

We next describe how to generate the four maximally entangled 2-qubit Bell states. We denote $|H>$ as the polarization of the x-ray electric field in the scattering plane (i.e., the plane containing the incident k-vector and the lattice k-vector $\vec{G}$), and  $|V>$  as the polarization orthogonal to the scattering plane. With the pump polarization in the scattering plane, the polarization of the emitted photon pair is
\begin{eqnarray}
\label{eq:currentinplane}
  |\psi>= \frac{1}{\sqrt{2}}\big[A(\theta_{p})|H_{s},H_{i}>+B(\theta_{p})|V_{s},V_{i}>\big]
    \end{eqnarray}
Here $\theta_{p}$, is the angle of the  pump k vector with regard to the atomic planes that are normal to $\vec{G}$. The coefficient $A(\theta_{p})$ is the nonlinear current density when the polarization of both signal and idler are in the scattering plane, and the coefficient $B(\theta_{p})$ is the current density when the polarization of both signal and idler are normal to the scattering plane. Similarly, when the pump polarization is normal to the scattering plane    \begin{eqnarray}
\label{eq:currentoutplane}
  |\phi>=\frac{1}{\sqrt{2}}\big[C(\theta_{p})|H_{s},V_{i}>+ D(\theta_{p})|V_{s},H_{i}>\big]
    \end{eqnarray}
The coefficient $C(\theta_{p})$ is the current density when the signal is polarized in the scattering plane and the idler is polarized normal to the scattering plane. The coefficient $D(\theta_{p})$ is the current density when the signal is polarized  normal to the scattering plane and the idler is polarized in the scattering plane. The quantities $A(\theta_{p}),B(\theta_{p}),C(\theta_{p})$, and $D(\theta_{p})$ are real functions of  $\theta_{p}$. This is different than conventional spontaneous parametric down conversion in the optical regime where the equivalent coefficients are generally complex \cite{Kwait,Mandel}.

The implication of Eq.~\eqref{eq:currentinplane} and~\eqref{eq:currentoutplane} is that the probabilities for generating the states $|H_{s},H_{i}>$  and $|V_{s},V_{i}>$ are   $|A(\theta_{p})|^{2}$ and  $|B(\theta_{p})|^{2}$ respectively; and the probabilities for generating the states $|H_{s},V_{i}>$  and $|V_{s},H_{i}>$ are $|C(\theta_{p})|^{2}$ and $|D(\theta_{p})|^{2}$. Consequently, unless either $|A(\theta_{p})|^{2}=0$ or $|B(\theta_{p})|^{2}=0$, Eq.~\eqref{eq:currentinplane} describes an entangled state. A maximally entangled state is obtained when $|A(\theta_{p})|=| B(\theta_{p})|$. Similarly, Eq.~\eqref{eq:currentoutplane} describes an entangled state which is maximally entangled when $|C(\theta_{p})|=|D(\theta_{p})|$. As we will show below, to produce all four of the Bell states, it is required that $\omega_s$ is not equal to $\omega_i$.

\begin{figure}[t]
\begin{center}
\includegraphics*[width=3.2 truein]{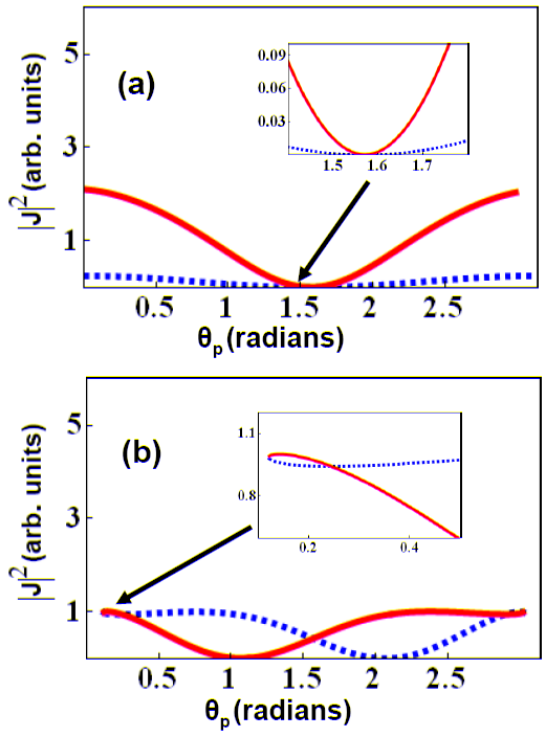}
\end{center}
\caption{ Polarization entanglement at degeneracy : (a) Square of the current density for the states $|H_{s},H_{i}>$ (solid green) and $|V_{s},V_{i}>$ ( dashed red), i.e.  $|A(\theta_{p})|^{2}$ and $|B(\theta_{p})|^{2}$ as function of the pump angle $\theta_p$.  (b) Square of the current density for the states $|H_{s},V_{i}>$ (solid green) and $|V_{s},H_{i}>$ (dashed red), i.e.  $|C(\theta_{p})|^{2}$ and $|D(\theta_{p})|^{2}$. The inset in (a) shows the intersection of the states $|H_{s},H_{i}>$ and $|V_{s},V_{i}>$. The inset in (b) shows the region containing the two intersection points on the left hand side of curves of the states $|H_{s},V_{i}>$ and $|V_{s},H_{i}>$. A similar region exists on the right hand side.}
\label{fig:degenerate}
\end{figure}
The procedure that we use to determine the Bell states is to first plot the square of the current density (Fig. 2) for each component of the Bell states. The intersections of the component curves determine the pump angles at which the magnitudes of the components of each Bell state are equal. We then, numerically, by simultaneous solution of the phase matching and current density equations, determine the sign of the components at the intersection. Consider a specific example: We choose diamond for the nonlinear medium and use the (111) lattice k-vector for phase matching. We take the  pump photon energy as 25 keV, and first consider the frequency-degenerate case where both the signal and idler energies are 12.5 keV. We solve the phase matching equations for  the angles of the signal $\theta_{s}$ and idler $\theta_{i}$ with regard to the atomic planes, and substitute the related electric fields into Eq. (2). With the polarization of the pump, signal, and idler chosen, the current density is a function of $\theta_{p}$ only. We calculate $|J_{s}|^{2}$ for each of the polarization states and plot the results in Fig.~\ref{fig:degenerate}.  Figures ~2(a)~and~2(b) show $|A(\theta_{p})|^{2}$ and $|B(\theta_{p})|^{2}$, and $|C(\theta_{p})|^{2}$ and $|D(\theta_{p})|^{2}$, respectively, all as function of $\theta_{p}$. From Fig.~\ref{fig:degenerate} (a) we see that  $|A(\theta_{p})|^{2}=|B(\theta_{p})|^{2}$ only when the nonlinearity is zero. Therefore at the degenerate frequency, and with the pump polarized in the scattering plane, the generated photon pairs are polarization entangled, but a pure Bell's state cannot be produced.

On the other hand, the solutions for $|C(\theta_{p})|^{2}=|D(\theta_{p})|^{2}$ are obtained at finite nonlinearity therefore allowing the Bell states $\frac{1}{\sqrt{2}}(|V_{s},H_{i}>\pm|H_{s},V_{i}>)$ to be generated. As shown in Fig.~\ref{fig:degenerate}(b), the equation $|C(\theta_{p})|^{2}=|D(\theta_{p})|^{2}$ has five solutions; i.e. there are five intersections between of the dashed and solid curves.  One solution is obtained at $\theta_{p}=\frac{\pi}{2}$ and corresponds to the state $\frac{1}{\sqrt{2}}(|V_{s},H_{i}>-|H_{s},V_{i}>)$. The other four solutions correspond to the state $\frac{1}{\sqrt{2}}(|V_{s},H_{i}>+|H_{s},V_{i}>)$. The Bell states and the corresponding angles of the pump, signal and idler are determined by numerically solving the phase matching and current density equations. The results are summarized in Table ~\ref{tab:degenerate}.\\
\begin{table}
\caption{\label{tab:degenerate}Angles of the k vectors of the pump, signal, and idler for producing maximally entangled states at the degenerate frequency.}
\begin{ruledtabular}
\begin{tabular}{llll}
 Bell's state &  $\theta_{p}$  & $\theta_{s}$ & $\theta_{i}$\\ \hline
  $(|H_{s},V_{i}>+|V_{s},H_{i}>)/\sqrt{2}$&0.1208&-0.1208&-0.1208\\
                                       &0.2434&-0.2434&0.2434\\
                                        &2.89819&2.89819&3.385\\
                                        &3.02079&3.26239&3.26239\\
 $(|H_{s},V_{i}>-|V_{s},H_{i}>)/\sqrt{2}$& $\frac{\pi}{2}$&2.2798&0.8617\\
\end{tabular}
\end{ruledtabular}
\end{table}
\begin{figure}[t]
\begin{center}
\includegraphics*[width=3.2 truein]{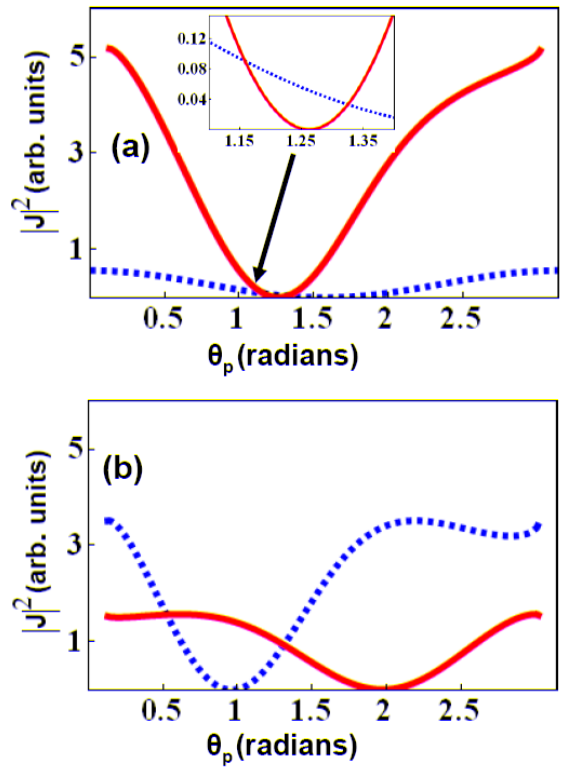}
\end{center}
\caption{Polarization entanglement at $20\%$ off of the degenerate frequency (a) Square of the current density for the states $|H_{s},H_{i}>$ (solid green) and $|V_{s},V_{i}>$ ( dashed red), i.e.  $|A(\theta_{p})|^{2}$ and $|B(\theta_{p})|^{2}$ as function of $\theta_p$.   (b) Square of the current density for the states $|H_{s},V_{i}>$ (solid green) and $|V_{s},H_{i}>$ (dashed red), i.e.  $|C(\theta_{p})|^{2}$ and $|D(\theta_{p})|^{2}$ as function of $\theta_p$ . The inset in (a) shows the intersection of the states $|H_{s},H_{i}>$ and $|V_{s},V_{i}>$. }
\label{fig:offdegenerate}
\end{figure}
Next, we analyze a case where the signal frequency is $20 \%$ off degeneracy as is shown in Fig.~\ref{fig:offdegenerate}. In this case it is possible generate each of the four Bell states. The corresponding angles of the pump, signal and idler are summarized in Table ~\ref{tab:offdegenerate}. We note that in contrast to the degenerate configuration, when off of degeneracy there is only one solution for each of the Bell states. That is, the angles of pump, signal and idler fields with regard to the atomic planes are uniquely defined by the Bell state.\\
\begin{table}
\caption{\label{tab:offdegenerate} Angles of the k vectors of the pump, signal, and the idler for producing maximally entangled states at $20\%$ off of the degenerate frequency.}
\begin{ruledtabular}
\begin{tabular}{llll}
Bell's state &  $\theta_{p}$  & $\theta_{s}$ & $\theta_{i}$\\ \hline
$(|H_{s},H_{i}>+|V_{s},V_{i}>)/\sqrt{2}$&1.3252 & 0.708018&2.13134 \\
$(|H_{s},H_{i}>-|V_{s},V_{i}>)/\sqrt{2}$&1.15798 & 0.512083&1.8803 \\
$(|H_{s},V_{i}>+|H_{s},V_{i}>)/\sqrt{2}$&0.525749&-0.0629786&0.84284\\
$(|H_{s},V_{i}>-|H_{s},V_{i}>)/\sqrt{2}$&1.31096&0.690752&2.11046\\
\end{tabular}
\end{ruledtabular}
\end{table}

To estimate the efficiency for the generation of Bell polarization states we solve the coupled Heisenberg-Langevin equations for each of the biphoton pairs, and numerically calculate the generation and coincidence count rates . For diamond, for the state
$(|H_{s},V_{i}>-|H_{s},V_{i}>)/\sqrt{2}$ with the signal frequency $20 \%$ above the degenerate frequency, a crystal length of 2 mm, a pump flux of $10^{13}$ photons/sec (available at the brightest synchrotron facilities), and detector apertures of
5 mrad $\times$ 5 mrad, the estimated coincidence  count rate is about 1 count per 15 seconds. This count rate is comparable to the count rate of previous experiments \cite{Adams}.

At x-ray wavelengths a polarizer may be constructed by  Bragg scattering with a Bragg angle of $\theta_{B}=45 ^{\circ}$.  When this is the case, the coefficient for scattering at $2 \theta_{B}$ is (ideally) zero  when polarized in the plane of incidence and unity when polarized perpendicular to the plane of incidence. Bragg polarizers with an energy bandwidth of $\Delta E/E \sim 10^{-2}$ might be constructed using mosaic crystals such as pyrolytic graphite \cite{Cross}.

In summary, this work has described a technique for using parametric down conversion at x-ray wavelengths to generate each of the four Bell polarization states. When off-degenerate this is done by choosing the angle of incidence of the pumping laser and polarizing it either in, or out, of the plane of incidence.  When at degeneracy, the pump must be polarized out of the plane of incidence, and only two of the four Bell states may be obtained.

The authors thank Jerry Hastings  for suggesting the x-ray polarizer, as above.  This work was supported by the U.S. Air Force Office of Scientific Research and the U.S. Army Research Office.

\end{document}